\documentclass{article}
\def\rf{\hspace{-1.5cm}}

\def\bsk{\bigskip}
\def\giantskip{\vskip2\bigskipamount}
\def\gsk{\giantskip}
\def\nin{\noindent}

\usepackage[scaled=0.92]{helvet}        
\usepackage{amsmath}
\usepackage{pstricks}           
\usepackage{graphicx}
\graphicspath{%
    {converted_graphics/}
    {//.PSF/.Mac/Volumes/Max's UNIX Home area/Under work/AA_Papers_in_progress/Exact_permutation/}
    {//.PSF/.Mac/Volumes/moldovan/Under work/AA_Papers_in_progress/Exact_permutation/}
    {/}
}
\usepackage[hang,flushmargin]{footmisc}
\begin{document}
\title{Exact conditional p-values from arbitrary ranking of a sample space: An application to genome-wide association studies}         
\author{Max Moldovan$^*$\; and \; Mette Langaas$^{**}$}        
\date{July 28, 2013}          
\maketitle
\nin$^*$ {\it Australian Institute of Health Innovation, University of New South Wales, Level 1 AGSM Building, Sydney NSW 2052, Australia; email: m.moldovan@unsw.edu.au} \\
$^{**}$ {\it  Department of Mathematical Sciences, Norwegian University of Science and Technology, N-7491 Trondheim, Norway; email: Mette.Langaas@math.ntnu.no}

\gsk
\centerline{SUMMARY}
\bsk
\nin We introduce a method for computation of exact conditional efficiency robust enumeration p-values for detection of genotype-phenotype associations at a single bi-allelic genetic locus. Our method can be based on any arbitrary ranking test statistics, such as efficiency robust test statistics or asymptotic p-values. The resulting p-values are exact conditional enumeration p-values and satisfy the basic statistical validity property $\Pr(P \leq \alpha | H_0) \leq \alpha$ for all parameters under the null hypothesis and all significance levels $\alpha$. Practically, the method allows performing statistically valid significance testing in genomic analyses with unknown modes of inheritance at individual bi-allelic genetic loci -- the situation typical in genome-wide association studies. We provide an open-source R code implementing the method. \\

\nin{\em Keywords:} mode of genetic inheritance; efficiency robust statistics; exact conditional inference; enumeration; genome-wide association study.

\vspace{0.35cm}
\setlength{\textheight}{575pt}
\baselineskip=20pt

\section{Introduction}

Genome-wide association studies (GWAS) consider hundreds of thousands of single nucleotide polymorphisms (SNPs) covering the entire human genome. Each SNP is normally represented by a bi-allelic locus and assessed for association with a specific genetic trait, usually in the context of a case--control study. Complex diseases such as asthma, diabetes and multiple sclerosis, among many others, are generally targeted by GWAS in order to identify common genetic variations as potential disease risk factors. The number of genetic markers in a particular GWAS can vary from several hundred thousands to several millions, depending on the platform used for genotyping and the type of genomes to be studied. For example, more SNPs are required for GWAS that utilize African populations than for GWAS involving European populations since linkage disequilibrium is much lower in the former. See Manolio (2010) for an interesting well illustrated introduction to GWAS.

At a given genetic locus, there is a pair of markers, called alleles, inherited from each of two parents.  Given a trait is passed through this locus, there are several modes of inheritance can be in effect. The dominant mode of inheritance requires the presence of a single `disease' allele from one of parents for a trait to be inherited. The recessive mode of inheritance requires `disease' alleles from both parents to be passed to the offspring for the trait to express. In case of the additive mode of inheritance, the trait is expressed only partly if a single `disease' allele is inherited, but express in full if both `disease' alleles are in place. There are a few more modes of inheritance can be specified depending on the degree to which the trait is expressed in an offspring, see Visscher {\it et al}. (2008) for the discussion of heritability concepts. 

When the mode of inheritance at a genetic locus is known, higher power of the test for genotype-phenotype association can be achieved through using a Cochran-Armitage trend test (CATT) under the explicit assumption of the specific genetic model, see Lettre {\it et al}. (2007) and Gonz\'{a}lez {\it et al}. (2008). In practice, however, the mode of inheritance is usually unknown. Under this typical scenario, the so-called efficiency robust tests (see Podgor {\it et al}., 1996) can be used -- the group of tests that remain sensitive to detection of genotype-phenotype associations even though the genetic model is either unknown or misspecified. 

There are several efficiency robust testing strategies. For example, the MAX test, first suggested by Freidlin {\it et al}. (2002), has been recommended by several authors, see Zheng and Gastwirth (2006) and Gonz\'{a}lez {\it et al}. (2008). This testing approach is implemented as a sequential application of several statistical tests optimal for alternative genetic models with retaining the most significant result. The traditional version of the MAX test, normally referred to as MAX3, is based on the three CATTs with scores motivated by dominant, recessive and additive genetic models. Alternatively, Person's chi-square test ($\chi^2$) can be included within the same MAX testing strategy, leading to MAX4, see Li {\it et al}. (2009). Zheng {\it et al}. (2009) demonstrated that $\chi^2$ test can be considered as a type of a trend test and also noted that this test is sensitive to detection of overdominant (underdominant) modes of inheritance. MIN2 is one more variant of the MAX test implemented as a combination of the additive CATT and $\chi^2$, see Joo {\it et al}. (2009). A slightly different efficiency robust testing strategy is known as MERT and is a weighted version of CATT optimal for recessive and dominant models, see Gastwirth (1985) and Freidlin {\it et al}. (2002). There are several more efficiency robust testing approaches can be specified and some authors even suggest applying a combination of different versions of efficiency robust tests within a single testing procedure, see Joo {\it et al}. (2009). 

In finite sample settings, many of the currently known and used efficiency robust tests are not guaranteed to lead to statistically valid inference. This is because the underlying computational procedures are based either on random sampling or on asymptotic distributions of efficiency robust statistics (Gonz\'{a}lez {\it et al}., 2008; Joo {\it et al}., 2010; So and Sham, 2011). For the methods that use simulated permutations (Sladek {\it et al}., 2007), the statistical inference will be valid, but a very high number of simulated permutations is needed to achieve the required precision for traditionally low GWAS type significance levels (often in the order of $10^{-8}$). Recently, Loley {\it et al}. (2013) attempted to unify the efficiency robust testing approaches by proposing a framework also leading to inference of unknown statistical validity. 

In the current paper, we introduce a computational procedure that takes as an input the ordering of a sample space imposed by any of test statistics or p-values, including the ones introduced above. The procedure outputs exact conditional enumeration p-values that satisfy the basic validity property $\Pr(P \leq \alpha | H_0) \leq \alpha$, for all parameters under the null hypothesis and all significance levels $\alpha$.

\section{Notation and the method}

Let the information on a single SNP be represented by the $2 \times 3$ contingency table given by Table~\ref{tab:table2x3-short}, where $x_i$ and $y_i$ are the counts of observed genotypes for $n_1$ cases and $n_2$ controls, respectively, with $n=n_1+n_2$. We denote this empirically observed table by $s^*(x_1,x_2|m_1,m_2,n_1,n_2)$, because all the other entries of the table can be calculated from these numbers. Note that for given $n_1$, $n_2$, $m_1$ and $m_2$, there is a finite number of possible contingency tables called a {\it reference set} (Verbeek, 1985) and denoted here by $S(m_1,m_2,n_1,n_2)$. Next let $T$ be an arbitrary ranking statistic with the value $t$ corresponding to the empirically observed table $s^*(x_1,x_2|m_1,m_2,n_1,n_2)$.  Given a general hypothesis of `$H_0:$ no association between genotypes and the case-control status of the subjects' tested against `$H_A:$ there is association between genotypes and the case-control status of the subjects', and larger values of $T$ being more hostile to the null $H_0$, the set of tables ranked lower or equal than the observed table $s^*(x_1,x_2|m_1,m_2,n_1,n_2)$ is given by the {\it critical set}:

\begin{equation}
R(x_1,x_2|m_1,m_2,n_1,n_2) := \{s(i,j|m_1,m_2,n_1,n_2): T \geq t \}.\\
\label{eq:critical-set}
\end{equation}

\begin{table}[htb]
\caption{Genotype counts at a bi-allelic locus.}
\begin{center}
\begin{tabular}{c| c c c| c} 
\hline
 & AA & AB & BB & Total\\ 
\hline
Case & $x_1$ & $x_2$ &  $x_{3}$ & $n_1$ \\
Control & $y_{1}$ & $y_{2}$ & $y_{3}$ & $n_2$ \\
 \hline
Total & $m_1$ & $m_2$ & $m_{3}$ & $n$ \\
\hline
\end{tabular}
\label{tab:table2x3-short}
\end{center}
\end{table}
\nin

\vspace{0.35cm}
By definition, a p-value is the probability of obtaining the outcome as extreme or worse than the empirically observed outcome $s^*(\cdot)$ under the null, which is just the probability of the critical set $R(\cdot)$. Under the null and based on the assumed underlying hypergeometric sampling scheme (see Lehmann (1986) for descriptions of alternative sampling schemes), the probability of obtaining each individual table $s(i,j|m_1,m_2,n_1,n_2)$ within the reference set can be computed as follows:

\begin{equation}
f(i,j|m_1,m_2,n_1,n_2) = \frac{ \binom {m_1}{i} \binom {m_2}{j} \binom {n-m_1-m_2}{n_1-i-j} } { \binom{n}{n_1} }. \\
\label{eq:hgd-2x3}
\end{equation}

\vspace{0.35cm}
\nin See Lloyd (1999) for the generalization of the central multivariate hypergeometric probability function given by (\ref{eq:hgd-2x3}). The p-value $p_{s^*,T}$ corresponding to $s^*(x_1,x_2|m_1,m_2,n_1,n_2)$ is the probability of the critical set $R(\cdot)$ given by (\ref{eq:critical-set}): 

\begin{equation}
p_{s^*,T}(x_1,x_2|m_1,m_2,n_1,n_2) = {\rm Pr}\Big(R(x_1,x_2|m_1,m_2,n_1,n_2)\Big) = \sum_{s \in R} f(i,j| m_1,m_2,n_1,n_2).\\
\label{eq:p-val-fce}
\end{equation}

\vspace{0.35cm}
\nin Note that $p_{s^*,T}$ is a Fisher-type conditional p-value by construction, inheriting positive (e.g. validity and empirical relevance) as well as negative (e.g. potential conservatism and computational challenges) aspects of Fisher's p-values.

\section{Numerical illustration}

Denote statistics obtained from CATTs optimal for dominant, recessive and additive models, respectively, by $T_D$, $T_R$ and $T_A$ (Sasieni, 1997, p.1258):

\vspace{-0.2cm}
$$T_D = \frac{n(n x_1 - n_1 m_1)^2}{n_1 m_1 (n - n_1) (n - m_1)}$$\\ 

\vspace{-1.5cm}
$$T_R = \frac{n (n x_3 - n_1 m_3)^2}{n_1 (n - n_1) (n m_3 - m_3^2)}$$\\

\vspace{-1.5cm}
$$T_A = \frac{n (n (x_2+2 x_3)-n_1 (m_2 + 2 m_3))^2}{n_1 (n - n_1) (n (m_2+4 m_3) - (m_2 + 2 m_3)^2)}$$\\


\vspace{-0.8cm}
\nin All three statistics asymptotically follow the chi-square distribution with one degree of freedom. The MAX3 test statistic is given by $T_{MAX3} = {\rm max} (T_D,T_R,T_A)$ with the observed value $t_{MAX3} = {\rm max} (t_D,t_R,t_A)$. For the empirically observed table $s^*(0,2|3,4,4,5)$, the p-value $p_{s^*,T_{MAX3}}$ can be computed as shown in Table~\ref{tab:example}. Specifically, there are 11 tables in the reference set $S(3,4,4,5)$ and only three tables in the critical set $R(0,2|3,4,4,5)$ given by (\ref{eq:critical-set}). Only the tables with the values of $T_{MAX3}$ statistics equally or more extreme then observed are included in the critical set, i.e. $T_{MAX3} \ge t_{MAX3}$. The resulted exact conditional efficiency robust p-value $p_{s^*,T}(0,2|3,4,4,5) = 0.0952$ and is the sum of $f(\cdot|m_1,m_2,n_1,n_2)$ given by (\ref{eq:hgd-2x3}) of the three tables in $R(0,2|3,4,4,5)$.

\begin{table}[htb]
\caption{The illustrative example is based on $(m_1,m_2,n_1,n_2) = (3,4,4,5)$ with an observed value $(x_1,x_2) = (0,2)$. The critical region $R$ is given by the lower part of the table under the horizontal line.}
\begin{center}
\label{tab:example}
\begin{tabular}{c c | c c c c | c}
\hline
$x_1$ & $x_2$ & $T_{D}$ & $T_{R}$ &    $T_{A}$ &   $T_{MAX3}$ &   $f(x_1,x_2|m_1,m_2)$ \\
\hline
1 &     2 & 0.2250 &   0.0321 &        0.1636 &        0.2250 &        {\gray 0.2857} \\
2 &     1 & 0.9000 &   0.0321 &        0.2557 &        0.9000 &        {\gray 0.1905} \\
1 &     3 & 0.2250 &   2.0571 &        0.2557 &        2.0571 &        {\gray 0.0952} \\
2 &     2 & 0.9000 &   2.0571 &        2.0045 &        2.0571 &        {\gray 0.1429} \\
1 &     1 & 0.2250 &   3.2143 &        1.7284 &        3.2143 &        {\gray 0.0952} \\
2 &     0 & 0.9000 &   3.2143 &        0.1636 &        3.2143 &        {\gray 0.0238} \\
0 &     3 & 3.6000 &   0.0321 &        1.7284 &        3.6000 &        {\gray 0.0635} \\
0 &     4 & 3.6000 &   2.0571 &        0.1636 &        3.6000 &        {\gray 0.0079} \\
\hline

{\bf 0} &    {\bf 2} & 3.6000 &   3.2143 &        4.9500 &       4.9500 &        0.0476 \\
3 &     0 & 5.6250 &    0.0321 &       2.0045 &        5.6250 &        0.0159 \\
3 &     1 & 5.6250 &   2.0571 &       5.4102 &         5.6250 &        0.0317 \\
\hline
                & & & & &                                       $p_{s^*,T} =$ &   {\bf 0.0952} \\
\hline
\end{tabular}
\end{center}
\end{table}

\section{Conclusion}

\nin The method we suggested above is by no means new. The initial idea can be traced back to Fisher (1935) and $P_{S,T}$ given by (\ref{eq:p-val-fce}) is based on the combinatorial results known for many decades, see Freeman and Halton (1951). Our contribution to the original Fisher's methodology is the idea of ordering the sample space, given by the reference set $S$, based on any arbitrary chosen ranking statistics, the efficiency robust test statistics in our case. We have borrowed this approach from the unconditional exact testing literature, see Barnard (1947) and Lloyd and Moldovan (2007) for the origination of the unconditional inference philosophy and one of the initial attempts to combine the conditional and unconditional types of exact inference, respectively.

To conclude, it should be pointed out that only the basic form of the adjustment procedure has been given above. In practice, more special cases can arise, such as the presence of covariates (e.g. additional SNPs, environmental factors or baseline factors) or involvement of additional shifted parameters (e.g. in power studies). While this is clearly the limitation of the presented procedure, the basic general exact conditional method introduced above gives a solid basis for further investigations to these and possibly several more theoretical and applied research directions. We provide an open-source R code to encourage and facilitate such investigations. The R code is available upon request from the authors.

\vspace{0.7cm}
\nin{\large \bf References}

\begin{quotation}



\rf Barnard, G.A. (1947) Significance tests for $2 \times 2$ tables. {\it Biometrika} {\bf 34}, 123-138.




\rf Fisher, R.A. (1935) The logic of inductive inference (with discussion). {\it Journal of the Royal Statistical Society} {\bf 98}, 39-54.

\rf Freeman, G.H. and Halton J.H. (1951) Note on an exact treatment of contingency, goodness of fit and other problems of significance. {\it Biometrika} {\bf 38}, 141-149.

\rf Freidlin, B., Zheng, G., Li, Z. and Gastwirth, J.L. (2002) Trend tests for case-control studies of genetic markers: Power, sample size and robustness. {\it Human Heredity} {\bf 53}, 146-152. 

\rf Gastwirth, J.L. (1985) The use of maximin efficiency robust tests in combining contingency tables and survival analysis. {\it Journal of the American Statistical Association} {\bf 80}, 380-384.

\rf Gonz\'{a}lez, J.R., Carrasco, J.L., Dudbridge, F., Armengol, L., Estivill, X. and Moreno, V. (2008) Maximizing association statistics over genetic models. {\it Genetic Epidemiology} {\bf 32}, 246-254.



\rf Joo, J., Kwak, M., Ahn, K. and Zheng, G. (2009) A robust genome-wide scan statistic of the Welcome Trust Case Control Consortium. {\it Biometrics} {\bf 65}, 1115-1122.


\rf Joo, J., Kwak, M. and Zheng, G. (2010) Improving power for testing genetic association in case-control studies by reducing the alternative space. {\it Biometrics} {\bf 66}, 266-276.

\rf Lehmann, E.L. (1986) {\it Testing statistical hypotheses}, 2nd ed., New York: Wiley.

\rf Lettre, G., Lange, C. and Hirschhorn, J.N. (2007) Genetic model testing and statistical power in population-based association studies of quantitative traits. {\it Genetic Epidemiology} {\bf 31}, 358-362.

\rf Li, Q., Zheng, G. and Yu, K. (2009) Robust tests for single-marker analysis in case-control genetic association studies. {\it Annals of Human Genetics} {\bf 73}, 245-252.

\rf Lloyd, C.J. (1999) {\it Statistical Analysis of Categorical Data}. New York: Wiley.

\rf Lloyd, C.J. and Moldovan, M. (2007) Unconditional efficient one-sided confidence limits for the odds ratio based on conditional likelihood. {\it Statistics in Medicine} {\bf 26}, 5136-5146.

\rf  Loley, C., K\"{o}nig, I.R., Hothorn, L. and Ziegler, A. (2013) A unifying framework for robust association testing, estimation, and genetic model selection using the generalized linear model. {\it  European Journal of Human Genetics}, to appear.

\rf Manolio, T.A. (2010) Genomewide association studies and assessment of the risk of disease. {\it New England Journal of Medicine} {\bf 363}, 166-176.








\rf Podgor, M.J, Gastwirth, J.L. and Mehta C.R (1996) Efficiency robust tests of independence in contingency tables with ordered classifications. {\it Statistics in Medicine} {\bf 15}, 2095-2105.


\rf Sasieni, P.D. (1997) From genotype to genes: doubling the sample size. {\it Biometrics} {\bf 53}, 1253-1261.

\rf So, H.C. and Sham, P.C. (2011) Robust association tests under different genetic models, allowing for binary or quantitative traits and covariates. {\it Behavior Genetics} {\bf 41}, 768-775.

\rf Sladek, R., Rocheleau, G., Rung, J., Dina, C., Shen, L., Serre, D., Boutin, P., Vincent, D., Belisle, A., Hadjadj, S., Balkau, B., Heude, B., Charpentier, G., Hudson, T.J., Montpetit, A., Pshezhetsky, A.V., Prentki, M., Posner, B.I., Balding, D.J., Meyre, D., Polychronakos, C. and Froguel, P. (2007) A genome-wide association study identifies novel risk loci for type 2 diabetes. {\it Nature} {\bf 445}, 881-885.


\rf Verbeek, A. (1985) A survey of algorithms for exact distributions of test statistics in $r \times c$ contingency tables with fixed margins. {\it Computational Statistics and Data Analysis} {\bf 3}, 159-185.

\rf Visscher, P.M., Hill, W.G. and Wray, N.R. (2008) Heritability in the genomics era -- concepts and misconceptions. {\it Nature Reviews Genetics} {\bf 9}, 255-266.



\rf Zheng, G. and Gastwirth, J.L. (2006) On estimation of the variance in Cochran-Armitage trend tests for genetic association using case-control studies.  {\it Statistics in Medicine} {\bf 25}, 3150-3159.

\rf Zheng, G., Joo, J. and Yang, Y. (2009) Pearson's test, trend test, and MAX are all trend tests with different types of scores. {\it Annals of Human Genetics} {\bf 73}, 133-140.


\end{quotation}

\end{document}